\documentclass[reqno,11pt]{amsart}
\usepackage{amsmath, latexsym, amsfonts, amssymb, amsthm, amscd}

\advance\hoffset -.75cm

\numberwithin{equation}{section}

\DeclareMathSymbol{\leqslant}{\mathalpha}{AMSa}{"36} 
\DeclareMathSymbol{\geqslant}{\mathalpha}{AMSa}{"3E} 
\DeclareMathSymbol{\eset}{\mathalpha}{AMSb}{"3F}     
\renewcommand{\leq}{\;\leqslant\;}                   
\renewcommand{\geq}{\;\geqslant\;}                   

\newcommand{\bra}{\langle}
\newcommand{\ket}{\rangle}

\newcommand{\cG}{\ensuremath{\mathcal G}}
\newcommand{\cH}{\ensuremath{\mathcal H}}
\newcommand{\cI}{\ensuremath{\mathcal I}}
\newcommand{\cJ}{\ensuremath{\mathcal J}}

\newcommand{\bbD}{{\ensuremath{\mathbb D}} }

\newcommand{\bbP}{{\ensuremath{\mathbb P}} }
\newcommand{\bbQ}{{\ensuremath{\mathbb Q}} }
\newcommand{\bbR}{{\ensuremath{\mathbb R}} }

\newcommand{\ga}{\alpha}

\newcommand{\gga}{\gamma}            
\newcommand{\gd}{\delta}
\newcommand{\gep}{\varepsilon}       
\newcommand{\gp}{\varphi}
\newcommand{\gr}{\rho}

\newcommand{\gP}{\Phi}

\newcommand{\gs}{\sigma}

\newcommand{\Fbound}{F^{\rm{res}}}

\begin{document}
\title{Current large deviations for Asymmetric Exclusion
Processes with open boundaries}

\author{T. Bodineau}
\address{Laboratoire de Probabilit{\'e}s et Mod{\`e}les Al{\'e}atoires,
CNRS-UMR 7599, Universit{\'e}s Paris VI $\&$ VII,
4 place Jussieu, Case 188, F-75252 Paris, Cedex 05}
\email{bodineau@math.jussieu.fr}

\author{B. Derrida}
\address{Laboratoire de Physique Statistique, Ecole Normale Sup{\'e}rieure,
24 rue Lhomond, 75231 Paris Cedex 05, France}
\email{derrida@lps.ens.fr}

\begin{abstract}
We study the   large deviation functional of the current
for the Weakly 
Asymmetric Simple Exclusion Process in contact with two reservoirs.
We compare
this  functional
in the large drift limit
to the one of the Totally Asymmetric Simple Exclusion Process, in
particular to 
the Jensen-Varadhan functional.
 Conjectures for generalizing the
Jensen-Varadhan functional to open systems are also stated.

\vspace{+0.2in}
{02.50.-r, 05.40.-a, 05.70 Ln, 82.20-w}
\end{abstract}

\today

\maketitle

\section{Introduction}

Recently, a theory  has been developed to describe the  current
fluctuations of diffusive stochastic gas in their non-equilibrium steady
state \cite{BDGJL2,BDGJL4,BDGJL5,BD,BD2,jsp, pjsb}.
The large deviation functional of the current can be obtained in
terms of a difficult variational problem; namely one needs to find the
most likely time dependent density profile conditionally to the value
of the current. 
So far some concrete informations have been derived: for example,
explicit expressions of the current cumulants have been derived 
in \cite{BD} and dynamical phase transitions have been predicted
in \cite{BDGJL4,BDGJL5,BD2}.
In the present paper, we analyze this functional 
in the case of a one-dimensional diffusive lattice gas of length
$N$, the simple exclusion process,  with a weak asymmetry $\nu/N$ (this
drift $\nu/N$ represents the effect of an external field acting on the
particles from left to right). We  will in particular consider the large
drift limit $\nu \to \infty$
 to investigate the relation
between the hydrodynamic large deviation functional of the Weakly Asymmetric 
Simple Exclusion Process (WASEP) and the one of the Totally Asymmetric 
Simple Exclusion Process (TASEP) which was computed by Jensen-Varadhan
\cite{JV}.

The density large deviation functional associated to the stationary measure of 
the TASEP in contact with reservoirs was computed in \cite{DLS1,DLS2}.
This computation was extended in \cite{ED} to the WASEP and the TASEP
functional was recovered  in the large drift limit
(at least for some range of the parameters). This convergence was not a
priori obvious as the hydrodynamic scalings in the WASEP and the TASEP are of
different nature:
for a system of size $N$, the time scales like $N^2$ in the WASEP and
like $N$ in the TASEP.
Our goal here is to show that a similar  convergence (from the WASEP to
the TASEP) holds for the current large deviations for an open system
in contact with two reservoirs 
(at least in some range of the values of the current).
A similar convergence was already observed in \cite{BD2} for the ring geometry.

In the present paper, we also discuss the crucial role 
played by the weak solutions of Burgers equation as in the Jensen-Varadhan 
theory \cite{JV}. 
It is well known that the weak solutions of  Burgers equation are not unique 
\cite{serre1}. 
A small viscosity enables to regularize the equation and to select the 
relevant weak solutions: from the PDE point of view, the others are 
disregarded as non-physical. 
These apparently non-physical weak solutions appear nevertheless as
the optimal way to produce some current fluctuations 
and we show that the regularization by a small viscosity remains valid
in the large deviation regime.
We stress that the correspondence between the WASEP and the TASEP is
obtained here at the macroscopic level (for the large deviation functionals).
We are in fact not able to make a precise derivation starting from the
microscopic models.

\medskip

The paper is organized as follows.
The WASEP in contact with reservoirs is introduced in Section
\ref{sec: WASEP}.
In  Section \ref{sec: WASEP current}, the current deviations 
of the WASEP are computed and in Section \ref{sec: From WASEP to TASEP},
they are  compared
 to the Jensen-Varadhan theory which is outlined
in Section \ref{sec: Jensen-Varadhan functional}.
The generalization of the Jensen-Varadhan functional to open systems
is also discussed in Section \ref{sec: From WASEP to TASEP}.

In the appendix we also calculate the probability of maintaining an
anti-shock in the viscous Burgers equation, and in the formal limit
$\nu=N$, we recover the Jensen-Varadhan functional \cite{JV}.

\section{Weakly Asymmetric Simple Exclusion Process}
\label{sec: WASEP}

In this section, we briefly review some known results on the hydrodynamic behavior
of the Weakly Asymmetric Simple Exclusion Process (WASEP).

	\subsection{The microscopic model}
	\label{subsec: WASEP}

We consider the WASEP on the chain $\bbD_N = \{ 1, \dots, N \}$ in contact 
with two reservoirs at the boundaries.
A particle configuration is given by $\eta \in \{0,1\}^N$, with 
$\eta_i = 1$ if a particle is at site $i$ and $\eta_i = 0$ otherwise.
The WASEP is a Markov process evolving according to the following
dynamical rules.
For a given $\nu>0$ and $N$ large enough, each particle attempts to jump to 
the right at rate $\frac{1}{2} - \frac{\nu}{2N}$ and to the left  at rate 
$\frac{1}{2} + \frac{\nu}{2N}$. 
If the target site is occupied, the jump is disregarded.

Particles are injected or removed only at the reservoirs, i.e. at
sites $1$ and $N$.
At the left boundary, i.e. at site $1$, particles are created with rate 
$c^+_1$ and annihilated with rate $c^-_1$. 
Similarly at the right boundary, particles are created at site $N$ with rate 
$c^+_N$ and annihilated with rate $c^-_N$. 

For any $N$, the expectation with respect to the invariant measure 
associated to the previous dynamics will be denoted by $\bra \cdot \ket_N$. 


\medskip

Under the action of the dynamics, a particle current flows through the system.
Let $Q(i,[s,s^\prime])$ be the integrated current through the bond
$(i,i+1)$ during the 
time interval $[s,s^\prime]$, i.e. the number of jumps from $i$ to $i+1$ minus 
the number of jumps from $i+1$ to $i$.
The space/time integrated current up to time $s$ will be defined as
\begin{eqnarray}
\label{eq: integrated current}
Q_s = \sum_{i=1}^{N-1} Q(i,[0,s]) \, .
\end{eqnarray}

\medskip

\noindent
{\bf Hydrostatic.}

We first consider the stationary regime.
The steady state density will be denoted by $\hat \gr$ and the 
steady state mean current by $\hat q$.
At the macroscopic level, the domain $\bbD_N$ is mapped into $\bbD = [0,1]$
and the local density of the steady state, at the macroscopic position
$x$, can be defined as 
\begin{eqnarray}
\label{eq: density static}
\forall x \in ]0,1[, \qquad
\hat \gr(x) = \lim_{\gep \to 0} \lim_{N \to \infty} \;
\left \bra \frac{1}{2 \gep N} \sum_{i=(x-\gep) N}^{(x+\gep )N} \eta_i \right \ket_N \, ,
\end{eqnarray}
where $\bra \cdot \ket_N$ is the invariant for the measure.
At the boundaries, $\hat \gr$ is given by the rates associated to the 
reservoirs
\begin{equation}
\label{eq: reservoirs}
\hat \gr(0) = \gr_a = \frac{c^+_1}{c^-_1 + c^+_1}
\quad \text{and} \quad
\hat \gr(1) = \gr_b = \frac{c^+_N}{c^-_N + c^+_N} 
\, .
\end{equation}
The local density satisfies Fick's law 
\begin{equation}
\label{eq: Fick}
\forall x \in ]0,1[, \qquad   
\frac{1}{2} \Delta_x \hat \gr - \nu \nabla_x \big( \gs( \hat \gr)
\big) = 0 \,, 
\end{equation}
with
\begin{equation}
\gs (\gr) = \gr (1 - \gr) \, .
\end{equation}

The hydrostatic equation \eqref{eq: Fick} has been derived for 
a broad class of stochastic models \cite{ELS1,ELS2,KOL,BEMM,LMS}
and in particular for the WASEP in \cite{ED}.

\bigskip

The mean current can also be deduced from \eqref{eq: Fick}
(see \cite{ELS1})
\begin{eqnarray}
\label{eq: mean current}
\hat q 
=   \lim_{N \to \infty} \; 
\lim_{s \to \infty} \left \bra \frac{Q_s}{s} \right \ket_N  
= - \frac{1}{2} \nabla_x \hat \gr + \nu \gs( \hat \gr) \,.
\end{eqnarray}
According to Fick's law the current through each bond is of the order $1/N$,
where $N$ is the length of the system.
The previous scaling comes from the fact that $Q_s$ is defined as the sum of 
the contributions of the $N$ bonds \eqref{eq: integrated current}.

The mean current can be explicitly computed from \eqref{eq: mean current}.
As an example, let us examine the case $\gr_a > \frac{1}{2} > \gr_b$
which becomes the {\it maximum current phase} in the TASEP,
that is in the limit of a strong asymmetry \cite{DEHP,SD,schutz,liggett}.
Equation \eqref{eq: mean current} implies that
\begin{eqnarray*}
2 = \int_{\rho_b}^{\rho_a} d \rho \; {1 \over {\hat q- \nu \gs (\rho)}}
=  \int_{v_b}^{v_a} d v \; {1 \over {\hat q- \frac{\nu}{4} + \nu v^2}} \, ,
\end{eqnarray*}
where we used the change of variables
$\gr \to 1/2 + v$ with $v_a = \gr_a -1/2, \  v_b = \gr_b -1/2$.
This leads to the following asymptotics for large $\nu$
\begin{eqnarray}
\label{eq: courant max}
\hat q = \frac{\nu}{4} +  \frac{\pi^2}{4 \nu} + 
o\left( \frac{1}{\nu} \right)\, . 
\end{eqnarray}

\medskip

\noindent
{\bf Hydrodynamic.}

We consider now the evolution of the stochastic process and denote by $\eta(s)$
the configuration  at time $s$.
Let $\gp$ be a macroscopic density profile. The evolution of the
particle system over the time interval $[0,s]$ and starting with 
initial data close to $\gp$ will be denoted by the probability
$\bbP^\gp_{N,s}$ (which depends on $\nu$ as well).
More precisely, $\bbP^\gp_{N,s}$ is  such that 
 the initial data are randomly chosen wrt the Bernoulli product
 measure with mean $\gp(i/N)$ at site $i$, then the time
evolution is given by the Markov dynamics.
To study the hydrodynamic limit, we introduce the  diffusive scaling 
\begin{eqnarray}
\label{eq: scaling 1}
x = \frac{i}{N}, \qquad  t = \frac{s}{N^2} \, .
\end{eqnarray}
At the macroscopic level, the particle system  is identified to the 
macroscopic density profile $\gr$ defined for any 
$(x,t) \in  \bbD \times [0,T]$ by
\begin{eqnarray}
\label{eq: density dynamics}
\gr(x,t) = \lim_{\gep \to 0} \lim_{N \to \infty} \; 
\frac{1}{2 \gep N} \sum_{i=(x-\gep) N}^{(x+\gep )N} \eta_i  (N^2 t)  \, .
\end{eqnarray}

If initially the discrete system is close (after rescaling) to the 
macroscopic profile $\gr(x,0) = \gp(x)$ then for any $t>0$ the rescaled
system remains close to the density profile solution of
\begin{equation}
\label{eq: WASEP diff}
\forall (x,t) \in \bbD  \times [0,T], \qquad 
\partial_t \gr = \frac{1}{2} \Delta \gr - \nu \nabla_x \big( \gs(\gr) \big) 
\quad \text{with} \quad 
\gr(0,t) = \gr_a, \ \gr(1,t) = \gr_b \, ,
\end{equation}
where $\gp$ is the initial data and $\gs(\gr) =  \gr (1-\gr)$.
This holds with a probability $\bbP^\gp_{N,T N^2}$ close to 1
\cite{ELS1,BEMM,KL,spohn}.

\bigskip

For any $(x,t) \in \bbD \times [0,T]$,
a local version of the space-time current can be defined as a space-time
average on the microscopic boxes
$\{ (x-\gep) N, (x+\gep )N \} \times [t N^2,(t+\gep)N^2]$ 
\begin{equation}
\label{eq: macro current}
j(x,t) = \lim_{\gep \to 0} \lim_{N \to \infty} \;
\frac{1}{2 \gep^2 N^2} \sum_{i=(x-\gep) N}^{(x+\gep )N} 
Q \big( i , [t N^2,(t+\gep)N^2]\big) \, .
\end{equation}
The scaling in \eqref{eq: macro current} can be understood as follows: 
$j$ is the rescaled current while the microscopic current through
each bond is $j/N$. Then summing over $2\gep N$ bonds during
a time interval $\gep N^2$ implies that one has to normalize by a 
factor $2 \gep^2 N^2$. 
The conservation of the number of particles at the microscopic level
becomes through \eqref{eq: density dynamics} and \eqref{eq: macro current}
\begin{equation*}
\partial_t \gr = - \nabla_x j \, .
\end{equation*}

	\subsection{Large deviations}
	\label{subsec: WASEP func}

The hydrodynamic equation \eqref{eq: WASEP diff} describes the typical 
behavior under the stochastic evolution. The probability of observing 
a different trajectory can also be computed.
In fact, one can even estimate the probability of observing a given
current of particles $j(x,t)$ and its corresponding density profile 
defined by
\begin{equation}
\label{eq: WASEP deviation}
\forall (x,t) \in \bbD  \times [0,T], \qquad 
\partial_t \gr = -  \nabla_x \, j \, , 
\end{equation}
with some initial data $\gr(x,0) = \gp(x)$ in $\bbD$
(Note that the current cannot be completely arbitrary: it should 
be such that the density \eqref{eq: WASEP deviation} remains in
$[0,1]$ and is equal to $\gr_a$ and $\gr_b$ at the boundaries at any time
$t>0$.)

\medskip

The probability of the event $\{ \eta \sim (j,\gr) \}$ that the
rescaled particle system has a current $j$ and remains close to 
the density profile $\gr$ introduced in 
\eqref{eq: WASEP deviation} is given asymptotically in $N$ by 
\begin{eqnarray}
\label{eq: LD asymptotics}
\bbP^\gp_{N,T N^2} \Big( \{ \eta \sim (j,\gr) \} \Big)
\approx \exp \Big( - N  \cI^\nu_{[0,T]}(j,\gr)  +  o( N ) \Big) \, ,
\end{eqnarray}
and the large deviation functional is  
\begin{eqnarray}
\label{eq: WASEP LD} 
\cI^\nu_{[0,T]}(j,\gr)   =  
\int_0^T  dt  \int_\bbD  dx \, 
\frac{\left( j(x,t) - \nu \gs \big( \gr(x,t) \big) + 
\frac{1}{2} \nabla_x  \gr(x,t) \right)^2}{2 \; \gs \big( \gr(x,t) \big)}  \, .
\end{eqnarray}
This can be interpreted as a local Gaussian fluctuation of the current
with a variance $\gs \big( \gr(x,t) \big)$ depending on the local density. 

\medskip

The hydrodynamic large deviation theory was originally introduced 
in \cite{KOV,spohn,KL} to estimate the probability of events
corresponding to atypical space/time density profiles.
Recent developments \cite{BDGJL4,BDGJL5,BD,jsp,pjsb} led to the  more general
expression \eqref{eq: WASEP LD} which enables one to control a deviation 
of the density as well as the associated current
(see the appendix of \cite{BD2} for a heuristic derivation).
Note that the current $j(x,t)$ appearing in 
(\ref{eq: LD asymptotics}\;--\;\ref{eq: WASEP LD})
is an average over a long microscopic time of
order $N^2$ (see \eqref{eq: macro current}).
Recent works \cite{DSt1,DSt2} have also
considered the correlations
between the density and the instantaneous current.
At present we do not see any direct connection between their
distribution of the instantaneous current and our $j(x,t)$.

\section{Current large deviations for a system with reservoirs}
\label{sec: WASEP current}

In this Section, we investigate the current large deviations for the
WASEP in contact with 2 reservoirs at densities $\gr_a$ and $\gr_b$.

\subsection{General estimates}

We are interested in the asymptotic probability of observing 
a deviation of the integrated current over a very long 
time interval $[0,T \, N^2]$, i.e. in the event that 
\begin{eqnarray}
\label{eq: current constraint} 
\frac{Q_{T\, N^2}}{T \, N^2}= \frac{1}{T} \int_{0}^T dt \int_0^1 dx \; j(x,t)
= \cJ 
\end{eqnarray}
for a very large macroscopic time $T$ (the correspondence between $Q$
and $j$ was stated in \eqref{eq: macro current}).
According to the large deviation principle \eqref{eq: LD asymptotics},
one has to minimize the functional \eqref{eq: WASEP LD}
over the time dependent density profiles under the 
mean current constraint \eqref{eq: current constraint}
\cite{BDGJL4,BDGJL5,BD2}. 
Suppose that the optimal way to achieve this deviation is 
by maintaining the density profile
close to some optimal time independent density profile.
Then, the variational problem simplifies and one has for $T$ large
\begin{eqnarray*}
\bbP^\gp_{N,T N^2} \Big( \frac{Q_{T\, N^2}}{T \, N^2} \sim \cJ \Big)
\approx \exp \Big( - N T G(\cJ) \Big) \, ,
\end{eqnarray*}
with
\begin{eqnarray}
\label{eq: WASEP LD reserv} 
G(\cJ)=  \inf_\gr \left\{ \int_0^1 \, dx \, 
\frac{\left(\cJ - \nu \gs(\gr(x)) + \frac{1}{2} \gr^\prime (x) \right)^2}{2 \;
\gs(\gr(x))} \right\} \, ,
\end{eqnarray}
where the infimum is taken over all the density profiles $\gr(x)$ with
boundary conditions $\gr_a$, $\gr_b$ prescribed by the reservoirs.
For simplicity, we wrote $ \gr^\prime  = \nabla_x \gr$.

This stationary regime assumption was already made in \cite{BD} when 
$\nu = 0$ and led
to an explicit prediction for all the cumulants of the integrated current 
$Q_T$ in diffusive lattice gases. 
It was however understood in \cite{BDGJL4,BDGJL5} that the stationarity 
assumption may not be satisfied for some diffusive lattice gases. 
In this case, a space/time dependent current is more favorable than
a constant current and $G(\cJ)$ does not give 
the correct order for observing the mean current 
$\cJ = \frac{1}{T} \int_{t = 0}^T \int_\bbD  j(x,t)$ (see
\cite{BDGJL4,BDGJL5,BD2}).

The expression \eqref{eq: WASEP LD reserv} 
can nevertheless be interpreted as the asymptotic
cost for observing a constant current $j(x,t) = \cJ$ for {\it all} 
times $t$ in $[0,T]$ (see \eqref{eq: macro current}).


\bigskip

Let us first
show that the optimal profile $\gr$ of the variational problem 
\eqref{eq: WASEP LD reserv} satisfies the relation 
\begin{eqnarray}
\label{eq: optimal profile}
\left( \frac{1}{2}  \gr^\prime  \right)^2 
= (\cJ - \nu \gs(\gr) \big)^2 + 2 K \gs(\gr) \, ,
\end{eqnarray}
where $K$ is an integration constant determined by the boundary conditions
$\gr_a,\gr_b$
(this formula is an extension of equation (15) obtained in \cite{BD}).
To see this, we write for a given density profile $\gr$
\begin{eqnarray}
\label{eq: 3.*}
\int_0^1 \, dx \, \frac{\left(\cJ - \nu \gs(\gr) 
+ \frac{1}{2}\gr^\prime \right)^2}{2 \;
\gs(\gr)}  
= 
\frac{1}{2} \int_0^1 \, dx \, \left( \cH(\gr) + 
\frac{(\gr^\prime)^2}{4 \gs(\gr)} 
\right)
- \int_{\gr_b}^{\gr_a} \, du \; \frac{ \cJ - \nu \gs(u)}{2\gs(u)} \, ,
\end{eqnarray}
where $\cH (\gr) = \frac{(\cJ - \nu \gs(\gr) \big)^2 }{\gs(\gr)}$.
The corresponding Euler equation is
\begin{eqnarray*}
\cH^\prime (\gr) -    \frac{\gr^{\prime \prime} }{2\gs(\gr)} + 
\frac{(\gr^\prime)^2 \gs^\prime (\gr)}{4\gs(\gr)^2} = 0 \, .
\end{eqnarray*}
Multiplying by $\gr^\prime$ we obtain
\begin{eqnarray*}
\nabla \cH (\gr) - \nabla \left( \frac{(\gr^\prime )^2}{4 \gs(\gr)}
\right) = 0 \, .
\end{eqnarray*}
This completes the derivation of \eqref{eq: optimal profile}.

\bigskip

In order to obtain more precise results, we are going now to specify the 
current deviation $\cJ$ as well as the boundary conditions.

		\subsection{Maximal current phase}
		\label{subsec: max current phase}

As an example, let us consider the case $\gr_a > 1/2 > \gr_b$, which 
becomes in the large $\nu$ limit the ``maximal current phase''.
According to \eqref{eq: courant max}, the mean current is of order $\nu/4$.
We are interested in the large $\nu$ asymptotics of the large deviation
functional $G(\nu q)$ (given in \eqref{eq: WASEP LD reserv}) for current 
deviations of the form 
$$\cJ = \nu q \, .$$

\bigskip

\noindent
{\bf Case 1 (current reduction): $q<1/4$.}

Let $\ga$ be the density for which the current is equal to $\nu q$ in
a stationary regime, we write $q = \gs(\ga) = \gs(1-\ga) <1/4$ 
(with $\ga > 1/2$).
As we will see the optimal profile to achieve this current deviation
concentrates close to $\ga$ or $1-\ga$ as $\nu$ tends to infinity.

Setting $K = \nu^2 k$, the equation \eqref{eq: optimal profile} of 
the minimizers of the functional $G(\nu q)$ can be rewritten
\begin{eqnarray}
\label{eq: minimizers asymptotic}
\left( \frac{1}{2 \nu} \gr^\prime \right)^2 = (q - \gs(\gr) \big)^2 
+ 2 k \gs(\gr) \, .
\end{eqnarray}

Different behaviors occur depending on the value of $q$, i.e. on the location
of $\ga$ wrt the reservoir densities $\gr_a$ and $\gr_b$.

\medskip

\noindent
$\bullet$ {\it Suppose that $\ga \in [\gr_b,\gr_a]$ or $1-\ga \in 
[\gr_b,\gr_a]$}.
As the optimal density profile goes smoothly from $\gr_a$ to $\gr_b$, 
it has to take the value $\ga$ (or $1-\ga$) and 
\eqref{eq: minimizers asymptotic} implies that $k \geq 0$. 
This implies that the optimal profile is decreasing. 
Integrating \eqref{eq: minimizers asymptotic} leads to 
\begin{eqnarray}
\label{eq: boundary conditions}
2 \nu = \int_{\rho_b}^{\rho_a} d \rho {1 \over \sqrt{(q- 
\sigma(\rho))^2 + 2 k \sigma(\rho)}} \, .
\end{eqnarray}
Thus as $\nu$ tends to infinity,  $k$ must vanish (in fact from
\eqref{eq: boundary conditions}, one can show that $k$ vanishes
exponentially fast wrt $\nu$). 
The optimal profile remains essentially equal to 
$\ga$ or $1-\ga$.

The expression \eqref{eq: minimizers asymptotic} of the optimal
profile combined with \eqref{eq: WASEP LD reserv} leads to
\begin{eqnarray} 
\label{eq: WASEP LD 2} 
G(\nu q)=  \nu \; \int_{\rho_b}^{\rho_a} d \rho {1 \over 2 \sigma(\rho) }
\left[ {(q- \sigma(\rho))^2 +  k \sigma(\rho) \over \sqrt{(q- 
\sigma(\rho))^2 + 2 k \sigma(\rho)}} - (q- \sigma(\rho)) \right]
\, .
\end{eqnarray}
As $k \to 0$ when $\nu \to \infty$
\begin{eqnarray} 
\label{eq: WASEP LD 3.0}
\lim_{k \to 0} {(q- \sigma(\rho))^2 +  k \sigma(\rho) \over \sqrt{(q- 
\sigma(\rho))^2 + 2 k \sigma(\rho)}} = 
\begin{cases}
\gs(\gr) - q, \qquad \text{if} \  \gr \in [1-\ga,\ga] \,, \\
 q - \gs(\gr), \qquad \text{otherwise} \, .
\end{cases} 
\end{eqnarray}
Let us define $\gr^+ = \min\{\ga,\gr_a\}$ and $\gr^- = \max \{1-\ga,\gr_b\}$,
then only the densities in $[\gr^-,\gr^+]$ contribute to the asymptotics
\begin{eqnarray}
\label{eq: WASEP LD 3}
\lim_{\nu \to \infty}  \frac{1}{\nu} G(\nu q)=   
\int_{\gr^-}^{\gr^+} d \rho \; {(\sigma(\rho) - q) \over \sigma(\rho)}
=   
(\gr^+ - \gr^-)- 
q \left[ \log \frac{\gr^+}{1-\gr^+} - \log \frac{\gr^-}{1-\gr^-}\right]   
\, .
\end{eqnarray}
where we used that $\gs(\gr) = \gr (1-\gr)$.


\bigskip

In particular, for $[1-\ga,\ga] \subset [\gr_b,\gr_a]$, this leads to 
the first main result of this paper
\begin{eqnarray}
\label{eq: JS static} 
\lim_{\nu \to \infty} \frac{1}{\nu} G(\nu q)= 
\big( g(1-\ga) - g(\ga) \big) \, , 
\end{eqnarray}
where 
\begin{eqnarray}
\label{eq: JS static 2} 
g(u) = u (1-u) \log \frac{u}{1-u} - u \, . 
\end{eqnarray}
In this case the optimal density profile is essentially a piecewise
constant function with three jumps over regions of width $1/\nu$:
a boundary layer near the left reservoir where the density varies 
from $\gr_a$ to $\ga$, another one from $1-\ga$ to $\gr_b$
near the right reservoir and a jump from $\ga$ to $1-\ga$.
The sharp discontinuity between $\ga$ and $1 - \ga$ will be
interpreted in Section \ref{subsec: Phys} as a convergence to an anti-shock
and the asymptotic cost \eqref{eq: JS static} coincides with the 
Jensen-Varadhan functional which will be defined in \eqref{eq: LD F}.

\bigskip

\noindent
$\bullet$ {\it Suppose that $[\gr_b,\gr_a] \subset [1-\ga,\ga]$}.

The optimal profile has to be close to $\ga$ or $1-\ga$ otherwise
$G(\nu q)$ would scale like $\nu^2$ (see \eqref{eq: 3.*}), thus the 
optimal profile cannot be always decreasing.
We first assume that it initially increases from $\gr_a$ to a local 
maximum  $\gga_\nu$ depending on $\nu$. 
According to \eqref{eq: minimizers asymptotic}, one has
\begin{eqnarray}
0  = (q - \gs(\gga_\nu) \big)^2 + 2 k \gs(\gga_\nu) \, .
\end{eqnarray}
The density cannot increase beyond $\gga_\nu$ otherwise 
the RHS of \eqref{eq: minimizers asymptotic} would be negative.
Thus the profile $\gr(x)$ increases 
from $\gr_a$ to $\gga_\nu$ in the interval $[0,x_\nu]$ and then  
decreases to $\gr_b$ in the interval $[x_\nu,1]$.
The analogous of \eqref{eq: boundary conditions} can be rewritten
as
\begin{eqnarray*}
2 \nu \; x_\nu = \int_{\rho_a}^{\gga_\nu} d \rho {1 \over \sqrt{(q- 
\sigma(\rho))^2 + 2 k \sigma(\rho)}} 
\qquad \text{and} \qquad
2 \nu \; (1-x_\nu) = \int_{\rho_b}^{\gga_\nu} d \rho {1 \over \sqrt{(q- 
\sigma(\rho))^2 + 2 k \sigma(\rho)}} 
\, .
\end{eqnarray*}
For large $\nu$, the important feature is the singularity at 
$\gga_\nu$ which implies that $\gga_\nu$ has to converge to $\ga$
and $k$ to 0 as $\nu$ diverges. 
In this case, the profile concentrates close to $\ga$.
Then \eqref{eq: 3.*} becomes
\begin{eqnarray*} 
&&  \nu \; \int_{\gr_a}^{\gga_\nu} d \rho {1 \over 2 \sigma(\rho) }
\left[ {(q- \sigma(\rho))^2 +  k \sigma(\rho) \over \sqrt{(q- 
\sigma(\rho))^2 + 2 k \sigma(\rho)}} + (q- \sigma(\rho)) \right]\\
&& \qquad \qquad   \qquad
+ \nu \; \int_{\rho_b}^{\gga_\nu} d \rho {1 \over 2 \sigma(\rho) }
\left[ {(q- \sigma(\rho))^2 +  k \sigma(\rho) \over \sqrt{(q- 
\sigma(\rho))^2 + 2 k \sigma(\rho)}} - (q- \sigma(\rho)) \right]
\, .
\end{eqnarray*}
In the limit $\nu \to \infty$, a computation similar to 
\eqref{eq: WASEP LD 3} applies: the contribution of the reservoir
$\gr_a$ vanishes and the asymptotic cost is
\begin{eqnarray}
\label{eq: Fbd b}
\int_{\gr_b}^{\ga} d \rho \; 
{(\sigma(\rho) - q) \over \sigma(\rho)}
= \ga - \gr_b + \ga(1-\ga) \left( \log \frac{\gr_b}{(1-\gr_b)} -
\log \frac{\ga }{(1-\ga)} \right)  \, . 
\end{eqnarray}

Similarly, if the profile first decreases from $\gr_a$ to $1-\ga$ and
then increases to $\gr_b$, then the asymptotic cost is 
\begin{eqnarray}
\label{eq: Fbd a}
\int_{1-\ga}^{\gr_a} d \rho \; 
{(\sigma(\rho) - q) \over \sigma(\rho)}
= \gr_a - (1- \ga) - \ga(1-\ga) \left( \log \frac{\gr_a}{(1-\gr_a)} -
\log \frac{(1-\ga) }{\ga} \right) \, .
\end{eqnarray}

\bigskip

Therefore, we can state our second main result.
If one defines the boundary cost by
\begin{eqnarray}
\label{eq: Fbd}
\forall u,v \in [0,1], \qquad
\Fbound (u,v)
= u - v - v(1-v) \left( \log \frac{u}{(1-u)} -
\log \frac{v }{(1-v)} \right) \, ,
\end{eqnarray}
then the optimal cost of the current deviation is given by 
\begin{eqnarray}
\label{eq: LD minimum}
\lim_{\nu \to \infty}  \frac{1}{\nu} G(\nu q)
=
 \min \big\{ \Fbound (\gr_a,1-\ga), - \Fbound (\gr_b,\ga) \big\}.
\end{eqnarray} 
where as before $q= \sigma(\alpha)= \alpha(1-\alpha)$.
We remark that the cost is due to the boundary effects and that the
contributions from the left and the right reservoirs decouple:
the first term in \eqref{eq: LD minimum} is a cost for a boundary
layer near the left reservoir with a density jump from $\gr_a$ to
$1-\ga$ whereas the second term is the cost for a boundary layer near the 
right reservoir with a density jump from  $\ga$ to $\gr_b$.
One can trigger a phase transition between the optimal profile 
close to $\ga$ and the one close to $1-\ga$ by varying $\gr_a$ and 
$\gr_b$ (the critical curve is $\gr_a = 1 -\gr_b$).
This provides an additional example of a non-equilibrium phase transition
associated to a current large deviation \cite{BDGJL4,BDGJL5,BD2}.

\bigskip

\noindent
{\bf Case 2  (current increase): $q>1/4$}.

Once again the optimal profile is determined by 
\eqref{eq: minimizers asymptotic}.
The optimal profile  decreases from $\gr_a$ to $\gr_b$
and equation \eqref{eq: boundary conditions} applies.
The minimum of the derivative $\gr^\prime$ 
(see \eqref{eq: minimizers asymptotic})
is reached at $\gr=1/2$, so that \eqref{eq: boundary conditions}
implies that for large $\nu$
\begin{eqnarray}
k = - 2 \left( q - \frac{1}{4} \right)^2  + \gep_\nu \, , 
\end{eqnarray}
where $\gep_\nu$ vanishes as $\nu$ tends to infinity.
The functional \eqref{eq: WASEP LD 2} can be rewritten, using 
\eqref{eq: boundary conditions}
\begin{eqnarray*} 
G(\nu q)= - \nu^2 k + \nu  \int_{\rho_b}^{\rho_a} d \rho {1 \over 2 \sigma(\rho) }
\left[ \sqrt{( q-  \sigma(\rho))^2 + 2 k \sigma(\rho)} - ( q-  \sigma(\rho)) \right]
\, .
\end{eqnarray*}
This leads to
\begin{eqnarray}
\label{eq: acceleration} 
G(\nu q) \sim   2 \nu^2 \left( q - \frac{1}{4} \right)^2
\qquad \text{and} \qquad
\lim_{\nu \to \infty} \frac{1}{\nu} G(\nu q) = \infty \, .
\end{eqnarray}
A deviation $q>1/4$ leads to asymptotics of $G(\nu q)$ in $\nu^2$ 
in contrast to a  deviation $q \in [0,1/4]$ where  $G(\nu q)$ 
scales like $\nu$.

\subsection{Boundary effects}
\label{subsec: boundary effects}

The current deviations select very specific density profiles. 
We are going now to estimate the large deviations associated 
to atypical density profiles instead of atypical currents.
The main motivation is to compute the boundary contribution for
a large drift $\nu$.

\medskip

We first fix $\gr_a > 1/2$ and let $\gr_b = \gga$. We are interested 
in evaluating the probability of maintaining a density profile close to $\gga$ 
over a very long time $T$, i.e.
\begin{eqnarray*}
\bbP^\gp_{N,T N^2}
\left(\forall t \in[0,T], \qquad
\int_0^1 dx \,  ( \gr(x,t) - \gga)^2 \leq \gep \right) 
\approx 
\exp \Big( - T N \cG (\nu,\gr_a, \gga) \Big) \, ,
\end{eqnarray*}
where the initial data $\gp$ is the profile uniformly equal to $\gga$.
The choice $\gr_b=\gga$ is made in order to suppress the boundary effect
of this reservoir so that the main contribution will depend only on
$\gr_a$.

As the constraint is time independent,
the large deviations \eqref{eq: LD asymptotics} imply that
\begin{eqnarray}
\label{eq: var  problem density} 
\cG (\nu,\gr_a, \gga) = \inf_{\cJ,\gr}
\left\{ \int_0^1  dx \, 
\frac{\left( \cJ - \nu \gs \big( \gr(x) \big) + 
\frac{1}{2} \gr^\prime (x) \right)^2}{2 \; \gs \big( \gr(x) \big)} 
\right\} \, ,
\end{eqnarray}
under the constraint that $\int_0^1 dx \,(\gr(x) - \gga)^2 \leq 
\gep$.

\bigskip

We are going to show that 
\begin{eqnarray}
\label{eq: LD boundary 1}
\forall \gr_a> \frac{1}{2}, \qquad 
\lim_{\gep \to 0}
\lim_{\nu \to \infty}  \frac{1}{\nu}  \cG (\nu,\gr_a,\gga )=  
\begin{cases}
\Fbound (\gr_a,\gga), \qquad  & \text{if $\gga \in [0,1-\gr_a]$}\\ 
g(\gga) - g(1-\gga) ,  \qquad & \text{if $\gga \in [1-\gr_a, 1/2]$} \\
0, \qquad & \text{if $\gga \in [1/2,1]$} 
\end{cases}
\end{eqnarray}
where $\Fbound$  was introduced in \eqref{eq: Fbd} and $g$ in 
\eqref{eq: JS static}.
This can be intuitively understood in view of the previous results on
the current large deviations.
$\Fbound (\gr_a,\gga)$ was computed in \eqref{eq: Fbd} as the
optimal cost for producing a current $\nu \gs(\gga)$ with an optimal 
profile close to $\gga < 1 -\gr_a$, thus the constraint
$\{ \int_0^1 dx \,(\gr(x) - \gga)^2 \leq \gep \}$ does not
change the asymptotic.
If $\gga \in [1-\gr_a,\gr_a]$ then the optimal profile to achieve a
current $\nu \gs(\gga)$
is concentrated close to $1-\gga$ and $\gga$ (see \eqref{eq: JS static}).
In this case, the cost is essentially given by the sharp jump between
$1-\gga$ and $\gga$. The location
of the discontinuity is not relevant to compute the asymptotic cost 
$g(\gga) - g(1-\gga)$. 
To satisfy the density constraint, it is enough to shift the discontinuity 
close to the reservoir $\gr_a$ so that all the energy becomes
concentrated in the boundary layer.

\medskip

We turn now to the derivation of \eqref{eq: LD boundary 1}.
The optimal current $\cJ$ (in the variational problem 
\eqref{eq: var  problem density}) is related to the optimal profile $\gr$ by
\begin{eqnarray}
\cJ \int_0^1  \frac{dx}{\gs \big( \gr(x) \big)} 
= 
\int_0^1  dx \, \frac{ \nu \gs \big( \gr(x) \big) - 
\frac{1}{2} \gr^\prime (x) }{ \gs \big( \gr(x) \big)} 
= \nu  + \int_\gga^{\gr_a}  \frac{d \gr }{2 \; \gs \big( \gr \big)} \, .
\end{eqnarray}
The density constraint implies that for $\nu$ large, the current is of
the order $\cJ = \nu q$, where $q \in I_\gd = [ \gs(\gga)-\gd,
\gs(\gga)+\gd]$ and $\gd$ vanishes as $\gep$ tends to 0.
Therefore, one can substitute the density constraint  by 
a  current constraint  and derive a lower bound for 
$\cG (\nu,\gr_a, \gga)$
\begin{eqnarray*}
\cG (\nu,\gr_a, \gga) \geq \inf_{q \in I_\gd} \;  
\inf_\gr \left\{ \int_0^1  dx \, 
\frac{\left( \nu q - \nu \gs \big( \gr(x) \big) + 
\frac{1}{2} \gr^\prime (x) \right)^2}{2 \; \gs \big( \gr(x) \big)} 
\right\} 
\geq \inf_{q \in I_\gd} G(\nu q) \, ,
\end{eqnarray*}
where $G$ was introduced in \eqref{eq: WASEP LD reserv}.
For $\gga \in [0,1/2]$, then the asymptotics of $G(\nu q)$
follow from \eqref{eq: JS static}, \eqref{eq: Fbd a}. 
When $\gd$ tends to 0, we deduce by continuity that the RHS of  
\eqref{eq: LD boundary 1} is a lower bound for the asymptotics of 
$\cG (\nu,\gr_a, \gga)$.
For $\gga \in [1/2,1]$, a cancellation similar to the one in 
\eqref{eq: WASEP LD 3} occurs and we recover 0 as a lower 
bound.

In order to derive an upper bound for $\cG (\nu,\gr_a, \gga)$, it is
enough to compute the asymptotic cost of well chosen density profiles.
Let us simply consider the case $\gga \in [1-\gr_a,\gr_a]$ and define
\begin{eqnarray}
\label{eq: optimal lagrange}
\left( \frac{1}{2 \nu}  \gr^\prime  \right)^2 
= \big( \gs(\gga) - \gs(\gr) \big)^2 + \gep^\prime (\gr - \gga)^2 +
2 k \gs(\gr) \, ,
\end{eqnarray}
where $k$ is chosen such that the boundary conditions are satisfied.
By tuning $\gep^\prime$ appropriately wrt $\gep$, we see that 
 this profile concentrates to the density $\gga$ for large $\nu$, so
that the density constraint is satisfied
\begin{eqnarray*}
\int_0^1  dx \, 
\frac{\left( \nu q - \nu \gs \big( \gr(x) \big) + 
\frac{1}{2} \gr^\prime (x) \right)^2}{2 \; \gs \big( \gr(x) \big)} 
\geq \cG (\nu,\gr_a, \gga)  \, .
\end{eqnarray*}
Proceeding as in \eqref{eq: WASEP LD 2}, \eqref{eq: WASEP LD 3}, 
we deduce that the asymptotic cost 
of the profile \eqref{eq: optimal lagrange} converges to 
$g(\gga) - g(1-\gga)$.
Similar arguments for the other values of $\gga$ enables to 
complete the proof of \eqref{eq: LD boundary 1}.

\bigskip

If $\gr_a<1/2$, then similar computations lead to 
\begin{eqnarray}
\label{eq: LD boundary 2}
\lim_{\gep \to 0}
\lim_{\nu \to \infty}  \frac{1}{\nu}  \cG (\nu,\gr_a,\gga )=  
\begin{cases}
\Fbound (\gr_a,\gga), \qquad  & \text{if $\gga \in [0,1/2]$} \\
\Fbound (\gr_a, 1-\gga), \qquad  & \text{if $\gga \in [1/2, 1-\gr_a]$} \\
0, \qquad  & \text{if $\gga \in [1-\gr_a, 1]$} \\
\end{cases}
\, , 
\end{eqnarray}
where $\Fbound$  was introduced in \eqref{eq: Fbd}.

The boundary cost of the right reservoir can be deduced by symmetries
(particle-hole,left-right).
For $\gr_b > 1/2$, the asymptotic cost of a discontinuity $\gga$ is
given by 
\begin{eqnarray}
\label{eq: LD boundary 3}
\lim_{\gep \to 0}
\lim_{\nu \to \infty}  \frac{1}{\nu}  \cG (\nu,\gga,\gr_b )=  
\begin{cases}
-\Fbound (\gr_b,\gga), \qquad  & \text{if $\gga \in [1/2,1]$} \\
-\Fbound (\gr_b, 1-\gga), \qquad  & \text{if $\gga \in [1-\gr_b,1/2]$} \\
0, \qquad  & \text{if $\gga \in [0,1-\gr_b]$} \\
\end{cases}
\, .
\end{eqnarray}
For $\gr_b < 1/2$
\begin{eqnarray}
\label{eq: LD boundary 4}
\lim_{\gep \to 0}
\lim_{\nu \to \infty}  \frac{1}{\nu}  \cG (\nu,\gga,\gr_b )=  
\begin{cases}
- \Fbound (\gr_b,\gga), \qquad  & \text{if $\gga \in [1-\gr_b,1]$}\\ 
g(1-\gga) - g(\gga) ,  \qquad & \text{if $\gga \in [1/2, 1-\gr_b]$} \\
0, \qquad & \text{if $\gga \in [0,1/2]$} 
\end{cases} \, .
\end{eqnarray}

\vskip1cm

In section \ref{subsec: max current phase}, we focused on the 
current large deviation for $\gr_a> 1/2 >\gr_b$. For other 
choices of the reservoir densities, similar computation would lead to
asymptotic costs which can be expressed also in terms of
\eqref{eq: LD boundary 1}, 
(\ref{eq: LD boundary 2}--\ref{eq: LD boundary 4}).
Further interpretation of the boundary effects is postponed until 
Subsection \ref{subsec: boundary}.


\section{Jensen-Varadhan functional}
\label{sec: Jensen-Varadhan functional}

In this Section, we recall some properties of the 
TASEP and discuss the hydrodynamic large deviations 
which were derived by Jensen and Varadhan \cite{JV}.
The connection between this functional and the previous results 
obtained for the WASEP will be detailed in Section 
\ref{sec: From WASEP to TASEP}.

	\subsection{TASEP}
	\label{subsec: TASEP}

Following the notation introduced for the WASEP,
we consider the TASEP on the domain $\bbD_N = \{ 1, \dots, N \}$ in contact 
with two reservoirs at the boundaries and denote 
a particle configuration at time $s$ by $\eta(s) \in \{0,1\}^N$.
For the TASEP, the particles only jump to the right with a constant
rate 1 if the neighboring site on their right is empty.
The reservoirs follow the same dynamical rules as the ones introduced 
in Section \ref{subsec: WASEP}.
Let $\bbQ^\gp_{N,s}$ be the probability measure associated to the TASEP 
on $\bbD_N$ during the microscopic time interval $[0,s]$ and starting from the 
Bernoulli product measure with local density at site $i$ equal to $\gp(i/N)$.

\medskip

The hydrodynamic behavior obeys a different  scaling from the one of the WASEP
\begin{eqnarray}
\label{eq: scaling 2}
 x = \frac{i}{N}, \qquad  t = \frac{s}{N} \, .
\end{eqnarray}
At the macroscopic level, the particle system is identified to the 
macroscopic density profile 
\begin{eqnarray*}
\gr(x,t) = \lim_{\gep \to 0} \lim_{N \to \infty} \  
\frac{1}{2 \gep N} \sum_{i=(x-\gep) N}^{(x+\gep )N} \eta_i  (N t)  \, .
\end{eqnarray*}
If initially the discrete system is close (after rescaling) to the 
macroscopic profile $\gr(x,0)$ then for any $t>0$, it was proven in 
\cite{Ba} (see also \cite{Re,KL}), that the rescaled
system remains close to the density profile evolving according to
Burgers equation
\begin{equation}
\label{eq: burgers}
\forall (x,t) \in  \bbD \times [0,T], \qquad \qquad 
\partial_t \gr + \nabla_x \big( \gs(\gr) \big) = 0
\qquad \text{with} \qquad \gs(\gr) = \gr(1-\gr) \, .
\end{equation}
For general initial conditions, 
the solution of this PDE is not well defined as shocks may occur 
after a finite time in the bulk and at the boundaries. 
For simplicity, we will not discuss the boundary effects and focus on the 
bulk singularities.  A detailed mathematical study of the PDE 
\eqref{eq: burgers} can be found in \cite{BLN,serre1,serre2}.
Notice also that the hydrodynamic limit can be derived beyond the 
shocks for a broad class of lattice gases \cite{Ba,Re}.

\bigskip

In order to take into account the shocks in the bulk, Burgers equation 
has to be rephrased in terms of  weak solutions.
Let $\gP$ be a smooth function with compact support strictly included  
in $\bbD \times [0,T]$, i.e. there is $\gd>0$ such that
\begin{equation*}
\forall x \not \in [\gd,1-\gd], \qquad \gP(x,t) = 0
\quad \text{and} \quad
\forall x \in \bbD, \qquad   \gP(x,0) = \gP(x,T)= 0 \, .
\end{equation*}
Integrating  \eqref{eq: burgers} by parts leads to 
\begin{equation}
\label{eq: weak burgers}
\int_0^T  dt \int_0^1 \, dx \, 
\big\{ \gr \, \partial_t \gP  +  \gs(\gr) \, \nabla_x \gP \big\} = 0 \, .
\end{equation}
A trajectory $\gr$ is said to be a {\it weak solution} of \eqref{eq: burgers} 
if it satisfies \eqref{eq: weak burgers} for any test function $\gP$
with compact support strictly included in $\bbD \times [0,T]$.
This formulation is not sufficient to prescribe the behavior
at the boundary, nevertheless we will be mainly interested in
bulk properties for which it is enough to consider \eqref{eq: weak burgers}.

For a given initial data, there is no uniqueness of the weak solutions.
Let us consider, as an example, a weak solution on $\bbR$.
We define a density profile with a discontinuity $a,b$ moving at speed $v$, i.e.
\begin{eqnarray}
\label{eq: shock}
\gr(x,t) =  \left\{
\begin{split}
a, \qquad \text{if} \quad x < vt \, , \\
b, \qquad \text{if} \quad  x > vt \, .
\end{split}
\right.
\end{eqnarray}
This profile is a solution of \eqref{eq: weak burgers} 
for any $a,b$ in $[0,1]$ provided that  the velocity
is given by
\begin{eqnarray}
\label{eq: speed}
v = \frac{\gs(a) - \gs(b)}{a-b} \, .
\end{eqnarray}
If $a<b$, the previous density profile (\ref{eq: shock})
is a physical solution. 
For $a>b$, the initial data with
the discontinuity $a,b$ at 0 evolves as a rarefaction fan so that 
the  previous density profile should be disregarded and considered
as a non physical event.
In the following, we will refer to the discontinuity $a<b$ as a {\em shock}
and  $a>b$ as an {\em anti-shock} \cite{serre1}.

\medskip

A way to select the physical solutions is to add a small viscosity
$\gep$ and to recover  Burgers equation in the limit $\gep \to 0$ 
\begin{equation}
\label{eq: burgers viscosity}
\forall (x,t) \in \bbD \times [0,T], \qquad 
\partial_t \gr + \nabla_x \big( \gs(\gr) \big)
= \frac{\gep}{2} \Delta \gr \, .
\end{equation}
The previous equation can be identified to the WASEP evolution
\eqref{eq: WASEP diff} after rescaling the time by a factor 
$\nu = 1/\gep$.
Thus in the hydrodynamic regime, the TASEP can be seen as the 
limit of the WASEP at large drift. Formally, this means that 
at the hydrodynamic level the limits $N \to \infty$ and 
$\nu \to \infty$ can be replaced by $\nu = N \to \infty$.

\subsection{Large deviations}
\label{sec: TASEP func}

The Jensen-Varadhan functional $\cJ_{[0,T]}$, introduced in \cite{JV},
provides an estimate of the probability that the rescaled  particle 
configuration remains close to a given density profile $\gr(x,t)$
over the macroscopic time interval $[0,T]$ 
\begin{eqnarray}
\label{eq: LD TASEP}
\bbQ^\gp_{N,TN} \Big( \{ \eta \sim \gr \} \Big)
\approx \exp \Big( - N  \cJ_{[0,T]}(\gr)  \Big) \, ,
\end{eqnarray}
with initial data $\gp(x) = \gr(x,0)$.
The key role of the weak solutions \eqref{eq: weak burgers} 
in the large deviation regime was understood in \cite{JV}: 
if $\gr$ is not a weak solution then $\cJ_{[0,T]}(\gr) = \infty$,
which means that the probability \eqref{eq: LD TASEP} vanishes
at an exponential order faster than $N$.
For general weak solutions, the Jensen-Varadhan functional
\cite{JV} is defined in terms of the Kruzkhov entropy 
\cite{serre1}.
In this paper, we will simply focus on a few concrete examples for which
an explicit representation of $\cJ_{[0,T]}$ can be stated.

\medskip


Let $\gr$ be a weak solution made of several shocks and anti-shocks 
(see \eqref{eq: shock})
and such that $\gr$ remains constant equal to $\gr_a$ and $\gr_b$ in a
small neighborhood of the reservoirs during the time interval $[0,T]$
(the latter condition enables us to disregard the boundary effects).
If $\gr$ has a single anti-shock of the type \eqref{eq: shock}
(i.e. $a>b$) during time $T$, then the Jensen-Varadhan functional
derived in \cite{JV} is given by 
$$\cJ_{[0,T]}(\gr) = T F(a,b)$$ 
with 
\begin{eqnarray}
\label{eq: LD F}
F(a,b) = g(b) - g(a) - \frac{\gs(b) - \gs(a)}{b-a} 
\big(h(b) - h(a) \big)   \, ,
\end{eqnarray}
where
\begin{eqnarray}
\label{eq: h,g}
\forall u \in [0,1], \qquad 
h(u) = u \log u +  (1-u) \log (1-u), \qquad
g(u) = u (1-u)\log \frac{u}{1-u} - u  \, .
\end{eqnarray}
(Remark that $h$ is the natural entropy associated to the
TASEP and that  $g'= h' \gs'$.) 
If there are several anti-shocks their contributions add up:
suppose that the evolution $\gr$ contains $\ell$ anti-shocks
$\{(a_i,b_i)\}_{i \leq \ell}$ each of them 
maintained during the
time interval $[t_i,s_i] \subset [0,T]$, then 
\begin{eqnarray}
\label{eq: JV sum}
\cJ_{[0,T]}(\gr) = \sum_{i =1}^\ell (s_i - t_i) F(a_i,b_i) \, .
\end{eqnarray}

\medskip

In the previous examples, the boundary effects are disregarded.
In fact, the Jensen-Varadhan theorem was derived on a torus 
and does not take into account the influence of the boundaries.
Nevertheless \eqref{eq: LD TASEP} should also apply in an open
system for weak solutions $\gr$ which are smooth at the boundary.
Heuristically, the Jensen-Varadhan functional is given by the local
cost of each anti-shock thus the regularity of the density at the 
boundary should be enough to decouple the bulk from the reservoirs.
The boundary contribution will be discussed in Section \ref{subsec: boundary}.

\section{From WASEP to TASEP}
\label{sec: From WASEP to TASEP}

\subsection{Physical relevance of the weak solutions.}
\label{subsec: Phys}

Some of the weak solutions of the Burgers equation (eg. the anti-shocks) are 
considered as unphysical and disregarded.
On the other hand, we showed  in \eqref{eq: JS static} that the anti-shocks 
with discontinuities $(\ga,1-\ga)$
arise naturally as the limit of the optimal density profiles which realize a
current large deviation $\nu \gs(\ga)$ when $\nu \to \infty$.
Furthermore, the large deviation functional associated to the 
WASEP converges to the Jensen-Varadhan functional corresponding 
to the stationary anti-shock with discontinuity $(\ga, 1-\ga)$
(see \eqref{eq: JS static} and \eqref{eq: LD F}). 
There are many ways of producing a current deviation 
even in a stationary regime, but conditionally to 
observing a current deviation of the form $\nu \gs(\ga)$,
the typical profiles are close to an anti-shock with discontinuity 
$(\ga, 1-\ga)$: the current deviation selects a specific weak solution.

Our observation 
shows that the viscosity regularization \eqref{eq: burgers viscosity}
remains valid beyond the law of large number regime
and that the Jensen-Varadhan functional can be understood as
a limit of the WASEP functional \eqref{eq: WASEP diff}.
Another example of an anti-shock traveling at non-zero velocity   
will be discussed in the Appendix.

\medskip

The correspondence between the WASEP and the TASEP process in 
the large deviation regime has been established at the macroscopic 
scale, i.e. for the large deviation functionals.
At the microscopic level, the correspondence 
is far from being obvious as the scaling limit for the WASEP 
\eqref{eq: scaling 1} and TASEP  \eqref{eq: scaling 2} are of different 
nature.
We can rewrite \eqref{eq: JS static} as 
\begin{eqnarray}
\label{eq: formel}
\bbP_{N,t}^\gp \left( \frac{Q_t}{t} \sim \nu q \right)
 \approx \exp \left( - t \left[ \frac{1}{N} G(\nu q) 
+  o\left(  \frac{1}{N} \right) \right]  \right) 
\approx \exp \left( -   \frac{t \nu}{N} F(\ga,1-\ga) \right)\, .
\end{eqnarray}
The order of the limits is the following: first $t \to \infty$, then
$N \to \infty$ and finally $\nu \to \infty$.
At least formally, in the limit $\nu = N \to \infty$, the asymptotics 
\eqref{eq: formel} is consistent with the Jensen-Varadhan theory.

Notice also that for the TASEP, the current distribution \cite{DL,DA} 
scales differently depending whether the current is decreased or 
increased.
In  \eqref{eq: acceleration}, a similar behavior
was pointed out for the WASEP. 
For $\nu$ large, the probability of observing a current deviation 
$\nu q$ with $q>1/4$ is given by
\begin{eqnarray}
\label{5.1'}
\bbP_{N,t}^\gp  \left( \frac{Q_t}{t} \sim \nu q \right)
\approx \exp \left( - t \left[ \frac{1}{N} G(\nu q) 
+  o\left(  \frac{1}{N} \right) \right]  \right) 
\approx \exp \left( -   \frac{t \nu^2}{N} \left( q - \frac{1}{4}
\right)^2 \right)\, .
\end{eqnarray}
At least formally, if $\nu = N$ the previous asymptotics 
is consistent with the scaling expected for the TASEP \cite{DL}: 
to increase the current beyond $1/4$ all the particles have to be accelerated.
One should not however expect the leading order in \eqref{5.1'} with
$\nu =N$ to be also valid for the TASEP. 
\eqref{5.1'} implicitly assumes that for large $N$, the measure
becomes locally a product measure.
We believe instead that in the TASEP, the optimal measure to increase the 
current remains different from a product measure even for large $N$.

	\subsection{Boundary contributions}
	\label{subsec: boundary}

The Jensen-Varadhan functional introduced in Section
\ref{sec: TASEP func} takes into account only the bulk 
large deviations and not the boundary contribution.
At present, there is no microscopic derivation of the
large deviations for the TASEP in contact with 
reservoirs.
For diffusive models, the large deviation functional of 
the system with reservoirs is simply the functional in the
bulk with the constraint that the densities at the reservoirs
are fixed equal to $\gr_a$ and $\gr_b$  (see \eqref{eq: WASEP LD}
or \cite{BDGJL3} for a rigorous derivation in the case
of the SSEP).
The Jensen-Varadhan functional cannot be generalized in this way,
as we know that even in the stationary regime, the reservoirs 
do not impose a fixed density:
the steady state of the TASEP can be discontinuous
at the boundary \cite{DEHP,krug,liggett,schutz,SD}.

In Section \ref{subsec: boundary effects}, we computed 
the probability of observing a sharp variation of the 
density at the boundary for large $\nu$.
As the Jensen-Varadhan functional can be interpreted as the limit of 
the WASEP large deviation functional in the bulk, we conjecture
that \eqref{eq: LD boundary 1}, 
(\ref{eq: LD boundary 2}--\ref{eq: LD boundary 4})
provide also the right expression for the boundary contribution.
If this is the case, the functional associated 
to the system with reservoirs should be the sum of the
Jensen-Varadhan functional (in the bulk) and of the
boundary effects. 

\bigskip

The previous conjecture cannot be validated without
a proper derivation from the microscopic model. Nevertheless, 
further comments can be made to justify our guess.
We will focus on the case $\gr_a>1/2$.
For the TASEP  on the half-line 
$\{1,2, \dots\}$, one can find, for any density $\gga$ in $[\frac{1}{2},1]$,
an invariant measure with mean asymptotically close to $\gga$
\cite{liggett}. 
This implies that there is no cost to maintain a discontinuity
$(\gr_a, \gga)$ for any $\gga$ in $[\frac{1}{2},1]$ as predicted
by \eqref{eq: LD boundary 1}.
For $\gga \in [1-\gr_a,\frac{1}{2}]$, \eqref{eq: LD boundary 1}
asserts that $1-\gga$ plays the role of an effective boundary.
The boundary layer is made of two pieces, first
a jump from $\gr_a$ to $1-\gga$ (which costs nothing as 
$1-\gga > 1/2$) and then an anti-shock with zero velocity between
$1-\gga$ and $\gga$ which has a cost given by the 
Jensen-Varadhan functional.
A similar interpretation holds for the expressions
(\ref{eq: LD boundary 2}--\ref{eq: LD boundary 4}).

\section{Conclusion}

In the present paper, we have analyzed the large deviation functional of
the current for the WASEP in contact with reservoirs. In the large drift
limit, we have obtained an expression \eqref{eq: JS static}
similar to the Jensen-Varadhan functional \eqref{eq: LD F}.
We have also obtained the cost \eqref{eq: LD boundary 1}, 
(\ref{eq: LD boundary 2}--\ref{eq: LD boundary 4}) of maintaining
boundary layers in the large drift limit and we believe that these
expressions should remain valid for the TASEP.
Of course, it would be necessary to validate our expressions from a
calculation starting from the microscopic model.
For general weak solutions, the Jensen-Varadhan functional has to be 
formulated in terms of the Kruzkhov entropy \cite{JV}. 
It would be interesting to relate the boundary costs \eqref{eq: LD boundary 1}, 
(\ref{eq: LD boundary 2}--\ref{eq: LD boundary 4})  to the more general
entropy condition introduced for open systems in \cite{BLN,serre2}.

A strong connection was derived in \cite{BDGJL2,FW}
between the hydrodynamic
large deviation functional and the one associated to the stationary measure: 
a density deviation in the stationary
measure can be interpreted as the optimal space/time cost to produce this
deviation starting from the steady state.
Thus, one could try to see whether the optimal trajectory to
generate a given steady state fluctuation   in the WASEP converges
to the optimal trajectory in the TASEP  (after all, if the large $\nu$
limit of the WASEP describes the  TASEP steady state at the level of
the density large deviation functional \cite{ED}, one might expect the large
drift limit to hold at the level of the whole trajectories).

Our calculation of the large deviations of the integrated current relies
on the hypothesis that the optimal density profiles are time-independent
and that the measure conditioned on a given current is in local
equilibrium. This led in particular to (\ref{5.1'}) which we do not expect to
hold in the TASEP: as for the ring geometry, the measure conditioned on
such an increase of the current should be significantly different from a
Bernoulli measure.

Even if we think that the current large deviations in the TASEP are
correctly given by \eqref{eq: WASEP LD 3}, \eqref{eq: JS static},
\eqref{eq: LD minimum} (i.e. the limit $\nu \to \infty$
of the WASEP), we believe that the distribution of the fluctuations of the
TASEP are not of diffusive nature and cannot be understood by taking
simply the large $\nu$ limit (for example in the WASEP the fluctuations
of density are Gaussian \cite{DELO} whereas in the TASEP they are not 
in the case of open boundary conditions \cite{DEL}).

\medskip

Finally,  it would be interesting to extend our approach 
 to particle systems with more general jump rates.
Hydrodynamic equations were derived for a broad class of microscopic 
models \cite{ELS1,KOL,KL}.
In this case, there exists
a density dependent diffusion coefficient $D(\rho)$ and a conductivity
coefficient $\gs(\rho)$ such that \eqref{eq: WASEP diff} becomes
\begin{equation}
\label{eq: WASEP diff general}
\forall (x,t) \in  \bbD \times [0,T], \qquad 
\partial_t \gr =  
\nabla_x( D( \gr) \nabla_x \gr) - \nu \nabla_x \big( \gs(\gr) \big) \, .
\end{equation}
The counterpart of the large deviation functional \eqref{eq: WASEP LD} 
can be expressed in terms of the coefficients $D$ and $\gs$
\cite{BDGJL4,BDGJL5,BD,BD2}
\begin{eqnarray}
\label{eq: WASEP LD general} 
\cI^\nu_{[0,T]}(j,\gr)   =  
\int_0^T  dt  \int_0^1  dx \, 
\frac{\left( j(x,t) - \nu \gs \big( \gr(x,t) \big) + D \big( \gr(x,t) \big) \nabla_x  \gr
(x,t) \right)^2}{2 \; \gs \big( \gr(x,t) \big)}  \, .
\end{eqnarray}
One can address the same type of questions as in this paper
and minimize this functional  under a current deviation 
constraint. The limiting expression (obtained for large drift $\nu$)
might provide the extension of the Jensen-Varadhan functional to 
more general asymmetric dynamics.

	\section{Appendix}
	\label{subsec: Traveling waves}

In Section \ref{subsec: max current phase}, a constant flux was imposed
and we proved the convergence of the large deviation functionals 
for $\nu$ large. 
We are now going to investigate the effect of a time
dependent current constraint.
More precisely, we are looking for the optimal way of producing a 
current with different values on both sides of a singularity evolving 
at velocity $v$ 
\begin{eqnarray}
\label{eq: asymptotic flux}
j(x,t) \sim  \left\{
\begin{split}
\nu \gs(a), \qquad \text{if} \quad x \ll vt , \\
\nu \gs(b), \qquad \text{if} \quad  x \gg vt \, .
\end{split}
\right.
\end{eqnarray}
For simplicity, we consider the hydrodynamic evolution in $\bbR$
instead of the finite system $\bbD=[0,1]$.
The large deviation functional  introduced in 
\eqref{eq: WASEP LD} can be defined in $\bbR$.
We would like to minimize the functional $\cI^\nu_{[0,T]}$ under 
the current constraint \eqref{eq: asymptotic flux} and 
find the optimal profile $\gr$ and the flux $j$
(both depending on $\nu$) for $\nu$ large.
By construction one has
\begin{eqnarray*}
\partial_t \gr = - \nabla j \, . 
\end{eqnarray*}
Furthermore, we restrict the class of density profiles to 
the traveling waves so that
\begin{eqnarray*}
\gr(x,t) = \gr(x- \nu v t), 
\qquad \text{and} \qquad
\nu v \gr^\prime =  j^\prime \, . 
\end{eqnarray*}
Integrating the previous equation, we deduce that $j = \nu (v \gr +c)$
and the constraint \eqref{eq: asymptotic flux} imposes that
\begin{eqnarray*}
v = \frac{\gs(a) - \gs(b)}{a-b}
\qquad \text{and} \qquad  
c = \frac{a \gs(b) - b \gs(a)}{a - b}
= - v a + \gs(a) = - v b + \gs(b) \, . 
\end{eqnarray*}
We remark that this imposes the same condition on the velocity 
as in \eqref{eq: speed}.

\vskip.5cm

The large deviation functional minimized over the traveling waves
is given by 
\begin{eqnarray}
\label{eq: WASEP LD dynamic} 
G(\nu,a,b) &=& \lim_{T \to \infty} \inf_{\gr} \left\{ \frac{1}{T} \;  \int_0^T \, dt \, 
\int_\bbR \, dx \, 
\frac{\big(j - \nu \gs(\gr) + \frac{1}{2} \gr^\prime \big)^2}{2 \;
\gs(\gr)} \right\}\\
&=& \inf_{\gr} \left\{ \int_\bbR \, dx \,  
\frac{\big( \nu (c+v \gr) - \nu \gs(\gr) + \frac{1}{2} \gr^\prime \big)^2}{2 \;
\gs(\gr)} \right\} \nonumber
\end{eqnarray}
where we used the explicit form of $j$ in terms of $\gr$.

\medskip

We  are going to recover the convergence to the Jensen-Varadhan functional
\eqref{eq: LD F}.
The optimal cost is given by the profile $\gr$ which minimizes 
\eqref{eq: WASEP LD dynamic}. We decompose $G$ into two terms 
\begin{eqnarray}
\label{eq: G1,G2} 
G_1 = \frac{1}{2}
\int_\bbR \, dx \, 
 \frac{\nu^2 \big(c+v \gr - \gs(\gr) \big)^2}{\gs(\gr)} 
+ \frac{(\gr^\prime)^2}{4\gs(\gr)}
\quad \text{and} \quad
G_2 =  \nu
\int_\bbR \, dx \, 
\frac{\big(c+v \gr - \gs(\gr) \big) \gr^\prime}{2\gs(\gr)}
\, .
\end{eqnarray}

The functional $G_2$ does not depend on the profile $\gr$
\begin{eqnarray*}
G_2 =  \frac{\nu}{2} \int_a^b \, d \gr \, 
\left(
\frac{c}{\gs(\gr)} -1 + v \frac{\gr}{\gs(\gr)}
\right) =
\frac{\nu}{2} \left[ c \log \frac{\gr}{1 -\gr} - \gr - v \log(1 -\gr)
\right]_a^b  \, .
\end{eqnarray*}
Using the explicit formula for $c$, we get
\begin{eqnarray*}
G_2 =  \frac{\nu}{2}
\left[ g(b) - g(a) - v \big( h(b) - h(a) \big)  \right]
= \frac{\nu}{2} F(a,b) \, .
\end{eqnarray*}

It remains to determine the optimal profile $\gr$ by minimizing $G_1$.
Following the same computation as in \eqref{eq: optimal profile},
we see that there is a constant $K$ such that
\begin{eqnarray*}
\left( \frac{1}{2} \gr^\prime \right)^2 = 
\nu^2  \big( c+v \gr - \gs(\gr) \big)^2 + 2 K \gs(\gr) \, .
\end{eqnarray*}
The constraint \eqref{eq: asymptotic flux} at infinity implies that $K =0$.
Thus we get
\begin{eqnarray*}
G_1  =  \frac{\nu}{2} \int_\bbR \, dx \,  
\frac{\big(c+v \gr - \gs(\gr) \big) \gr^\prime}{\gs(\gr)} 
= G_2 \, .
\end{eqnarray*}

Combining the previous results, we get
\begin{eqnarray*}
G (\nu q) =  
\begin{cases}
\nu F(a,b), \qquad & \text{if} \quad a>b \\  
0, & \text{if} \quad a<b
\end{cases} 
\, .
\end{eqnarray*}
For $\nu$ large (and in fact for all $\nu$), 
$G (\nu q)/\nu$ is equal to the Jensen-Varadhan functional
\eqref{eq: LD TASEP}, \eqref{eq: LD F}.
As expected the shocks ($a<b$) behave differently from the 
anti-shocks ($a>b$).
The optimal profile has a sharp slope in an interval of size $1/\nu$ 
centered around $v t$ and the cost to maintain the anti-shock
is essentially located in this region.
This confirms the Jensen-Varadhan theory which asserts that 
the contribution of each anti-shocks decouple \eqref{eq: JV sum}.
For this reason the previous approximation procedure should be
also valid for a more general deviation of the current with 
several discontinuities: in the limit $\nu \to \infty$, 
each anti-shock should contribute
independently to the large deviation functional.

In \cite{BD2}, it was argued that some current deviations for 
the WASEP on a ring are achieved by a traveling wave. 
In the large drift limit, the exponential cost of these traveling 
waves was also related to the  Jensen-Varadhan functional.

\bigskip

\noindent
{\it Acknowledgments.}
We  are very grateful to  C. Bahadoran for explaining us the 
Jensen-Varadhan theory and for very useful discussions.

\end{document}